\documentclass{article}
\usepackage{spconf,amsmath,graphicx}

\usepackage{amsmath,amsfonts}
\usepackage{amssymb}
\usepackage{algorithm}
\usepackage{array}
\usepackage[caption=false,font=normalsize,labelfont=sf,textfont=sf]{subfig}
\usepackage{textcomp}
\usepackage{stfloats}
\usepackage{url}
\usepackage{verbatim}
\usepackage{graphicx}
\usepackage{cite}

\usepackage[acronym]{glossaries}
\newacronym{ict}{ICT}{information and communications technology}
\newacronym[plural=BSs,firstplural=base stations (BSs)]{bs}{BS}{base station}
\newacronym[plural=PAs,firstplural=power amplifiers (PAs)]{pa}{PA}{power amplifier}
\newacronym{sdr}{SDR}{signal-to-distortion ratio}
\newacronym{snr}{SNR}{signal-to-noise ratio}
\newacronym{sinr}{SINR}{signal-to-interference-plus-noise ratio}
\newacronym{sndr}{SNDR}{signal-to-noise-and-distortion ratio}
\newacronym{snidr}{SNIDR}{signal-to-noise-and-interference-and-distortion ratio}
\newacronym{los}{LOS}{line-of-sight}
\newacronym{z3ro}{Z3RO}{zero third-order distortion}
\newacronym{mimo}{MIMO}{multiple-input multiple-output}
\newacronym{siso}{SISO}{single-input single-output}
\newacronym{mrt}{MRT}{maximum ratio transmission}
\newacronym{dpd}{DPD}{digital pre-distortion}
\newacronym{ula}{ULA}{uniform linear array}
\newacronym{se}{SE}{spectral efficiency}
\newacronym{ee}{EE}{energy efficiency}
\newacronym{tdma}{TDMA}{time-division multiple access}
\newacronym{ran}{RAN}{radio access network}
\newacronym{lte}{LTE}{Long-Term Evolution}
\newacronym{nr}{NR}{New Radio}
\newacronym{sdma}{SDMA}{space-division multiple access}
\newacronym{csi}{CSI}{channel state information}
\newacronym{psd}{PSD}{power spectral density}
\newacronym{aclr}{ACLR}{adjacent channel leakage ratio}
\newacronym{cpu}{CPU}{central processing unit}
\newacronym{iid}{i.i.d.}{independent and identically distributed}
\newacronym{ue}{UE}{user equipment}
\newacronym{nlos}{NLoS}{non-line-of-sight}
\newacronym{hpbm}{HPBM}{half power beam width}
\newacronym{rts}{RTS}{ray tracing simulator}
\newacronym{ag}{AG}{array gain}
\newacronym{arp}{ARP}{Antenna Reference Point}
\newacronym{ecdf}{eCDF}{Empirical Cumulative Distribution Function}
\newacronym[plural=CDFs,firstplural=cumulative distribution functions (CDFs)]{cdf}{CDF}{cumulative distribution function}
\newacronym{evm}{EVM}{Error Vector Magnitude}
\newacronym{fft}{FFT}{fast Fourier transform}
\newacronym{ifft}{IFFT}{inverse fast Fourier transform}
\newacronym{ib}{IB}{in-band}
\newacronym{im}{IM}{intermodulation}
\newacronym{mf}{MF}{matched filtering}
\newacronym{mr}{MR}{maximum ratio}
\newacronym{ofdm}{OFDM}{orthogonal frequency-division multiplexing}
\newacronym{oob}{OOB}{out-of-band}
\newacronym{papr}{PAPR}{peak-to-average power ratio}
\newacronym{trp}{TRP}{Transmission Reception Point}
\newacronym{zf}{ZF}{zero-forcing}
\newacronym{dl}{DL}{downlink}
\newacronym{ul}{UL}{uplink}
\newacronym{rzf}{RZF}{regularized zero forcing}
\newacronym{slc}{SLC}{spatial leakage suppression}
\newacronym{kpi}{KPI}{key performance indicator}
\newacronym{awgn}{AWGN}{additive white gaussian noise}
\newacronym{qam}{QAM}{quadrature amplitude modulation}
\newacronym{amam}{AM/AM}{amplitude modulation to amplitude modulation}
\newacronym{ampm}{AM/PM}{amplitude modulation to phase modulation}
\newacronym{zmcscg}{ZMCSCG}{zero mean circularly symmetric complex Gaussian}
\newacronym[plural=FOCs,firstplural=first-order conditions (FOCs)]{foc}{FOC}{first-order condition}
\newacronym{3gpp}{3GPP}{3rd Generation Partnership Project}
\newacronym{6G}{6G}{sixth-generation}
\newacronym{5G}{5G}{fifth-generation}
\newacronym{4G}{4G}{fourth-generation}
\newacronym{qos}{QoS}{quality of service}
\newacronym{afe}{AFE}{analog front-end}
\newacronym{dbb}{DBB}{digital baseband}
\newacronym{bb}{BB}{baseband}
\newacronym{ru}{RU}{radio unit}
\newacronym{rru}{RRU}{remote radio unit}
\newacronym{aau}{AAU}{active antenna unit}
\newacronym{fdd}{FDD}{frequency-division duplexing}
\newacronym{tdd}{TDD}{time-division duplexing}
\newacronym[plural=CQIs]{cqi}{CQI}{channel quality indicator}
\newacronym{dtx}{DTX}{discontinuous transmission}
\newacronym{mudtx}{\textmu DTX}{micro-discontinuous transmission}
\newacronym{drx}{DRX}{discontinuous reception}
\newacronym{fpga}{FPGA}{field-programmable gate array}
\newacronym{asic}{ASIC}{application-specific integrated circuit}
\newacronym{rfsoc}{RFSoC}{RF system-on-chip}
\newacronym{tx}{TX}{transmitting}
\newacronym{rx}{RX}{receiving}
\newacronym{rf}{RF}{radio-frequency}
\newacronym{mmwave}{mmWave}{millimeter wave}
\newacronym{ldpc}{LDPC}{low-density parity-check}
\newacronym{gops}{GOPS}{Giga operations per second}
\newacronym{FR3}{FR3}{Frequency Range 3}
\newacronym[plural=LNAs,firstplural=low-noise amplifier (LNAs)]{lna}{LNA}{low-noise amplifier}
\newacronym[plural=DACs,firstplural=digital-to-analog converters (DACs)]{dac}{DAC}{digital-to-analog converter}
\newacronym[plural=ADCs,firstplural=analog-to-digital converter (ADCs)]{adc}{ADC}{analog-to-digital converter}
\newacronym{gan}{GaN}{gallium nitride}
\newacronym{cmos}{CMOS}{complementary metal-oxide-semiconductor}
\newacronym{gaas}{GaAs}{gallium arsenide}
\newacronym{sige}{SiGe}{silicon-germanium}
\newacronym{pae}{PAE}{power-added efficiency}
\newacronym{lo}{LO}{local oscillator}

\glsdisablehyper
\usepackage{xcolor}
\usepackage{hyperref}
\usepackage{pgfplots}
\pgfplotsset{compat=1.18}
\usepgfplotslibrary{fillbetween}
\usepackage{algpseudocode}
\usepackage{balance}
\usepackage{pstricks}
\usepackage{enumitem}
\usepackage{fontawesome5}

\makeatletter
\newcommand{\github}[1]{%
   \href{#1}{\faGithubSquare}%
}
\makeatother

\usetikzlibrary{shapes,spy,circuits,arrows,patterns,decorations.pathreplacing,arrows.meta,positioning}
\usetikzlibrary{pgfplots.fillbetween}

\newcommand{\sublow}[2]{{#1}_\mathsf{#2}}
\newcommand{\subupp}[2]{{#1}_\mathsf{\scriptscriptstyle{#2}}}
\newcommand{\Mant}{\sublow{M}{ant}}
\newcommand{\MRF}{\subupp{M}{RF}}
\newcommand{\Pa}{\sublow{P}{a}}
\newcommand{\Pcons}{\sublow{P}{cons}}
\newcommand{\Pmax}{\sublow{P}{max}}
\newcommand{\vect}[1]{\boldsymbol{\mathrm{#1}}}
\newcommand{\mat}[1]{\boldsymbol{\mathrm{#1}}}
\newcommand{\tp}[1]{#1^{\scriptscriptstyle\mathsf{T}}}
\newcommand{\hr}[1]{#1^{{\scriptscriptstyle\mathsf{H}}}}
\newcommand\abs[1]{\left|{#1}\right|}
\newcommand{\norm}[1]{\left\lVert#1\right\rVert}
\newcommand*{\inC}[1]{\in\mathbb{C}^{#1}}
\newcommand{\expt}[1]{\mathbb{E}\left\{#1\right\}}

\title{A Parametric Power Model of Upper Mid-Band (FR3) Base Stations for 6G}
\name{Emanuele Peschiera$^\star$, Sangbu Yun$^\dagger$,
Youngjoo Lee$^\ddagger$, Liesbet Van der Perre$^\star$, and
Fran\c{c}ois Rottenberg$^\star$
\vspace*{-.2cm}
\thanks{$\!\!$This work has received funding from FWO (Travel Grant no.~V434924N).}}
\address{$^\star$ESAT-DRAMCO, KU Leuven, Ghent, Belgium,
$^\dagger$Department of EE, POSTECH,\\ Pohang, Republic of Korea, 
$^\ddagger$School of EE, KAIST, Daejeon, Republic of Korea
\vspace*{-.2cm}}

\begin{document}
\ninept

\maketitle

\begin{abstract}
Increasing attention is given to the upper mid-band or \gls{FR3}, from $7$ to $24$ GHz, 
in the research towards \gls{6G} networks. Promises of offering large data rates at 
favorable propagation conditions are leading to novel \gls{FR3} \gls{bs} architectures,
with up to thousands of antenna elements and \gls{rf} chains.
This work investigates the power consumption of prospective \gls{FR3} \glspl{bs}
and its relation to the delivered data rates. We model the power consumed by
digital and analog signal processing, \glspl{pa}, and supply and cooling during four phases
(data, signaling, micro-sleep, and idle) in downlink and uplink. 
Hybrid partially-connected beamforming is compared to fully-digital one. 
Results show that, for \gls{bs} arrays with $1024$ antennas at $30\%$ of load, 
the \gls{pa} consumes most of the power when $64$ or less RF chains are utilized, while
the digital and analog processing consumption takes over when the number of RF chains is 
$512$ or more. The digital plus analog processing consumes $2\times$ to $4\times$ 
more than the \gls{pa} for fully-digital beamforming. Hybrid beamforming achieves 
$1.3$ Gbit/s/user in downlink while improving the energy efficiency by 
$1.4\times$ compared to fully-digital beamforming.
\end{abstract}

\begin{keywords}
Power consumption, 6G mobile communication, Base stations, MIMO systems, X-band.
\end{keywords}

\glsresetall
\section{Introduction}

The upper mid-band, roughly ranging from 7 to 24 GHz, is one of the main candidate frequency bands
to provide a $10\times$ capacity increase in \gls{6G} networks~\cite{Andrews_2024}. 
This band is expected to trade-off the benefits of sub-6 GHz bands in terms of propagation 
and coverage, with the large bandwidths offered by \gls{mmwave} frequencies~\cite{Kang_2024}. 
Within the \gls{3gpp}, the upper mid-band is referred to as \gls{FR3}.  
Among recent works, considerations on channel propagation at \gls{FR3} are provided in~\cite{Zhang_2025}, 
while indoor channel measurements are presented in~\cite{Shakya_2024}. 
Prototyping and channel sounders are discussed in~\cite{Mezzavilla_2025}.
The work~\cite{Bjornson_2025} examines the transition to fully-digital \gls{mimo}
arrays having more antenna elements than in current massive \gls{mimo}. 

Evaluating the \gls{bs} power consumption is a fundamental task to link communication 
performance with environmental and economic sustainability, with the topic being
widely investigated for sub-6 GHz \glspl{bs}~\cite{Busch_2024}. 
A recent work has proposed a parametric model of sub-6 GHz \glspl{bs} using
on-site measurements and equipment documentation~\cite{Golard_2024}. For \gls{mmwave}
\glspl{bs}, the model~\cite{Desset_2020} quantifies the consumption of \gls{pa},
analog, and digital components, while works as~\cite{Ribeiro_2018} focus on 
analog and \gls{pa} consumption versus achieved spectral efficiency by using hybrid beamforming.
The capacity and power consumption of heterogenous \gls{bs} deployments with different frequency 
bands including \gls{FR3} are analyzed in~\cite{Lopez-Perez_2024}.

In this work, inspired by~\cite{Golard_2024,Desset_2020} that considered FR1 and FR2 BSs, 
we propose the first detailed power consumption model for \gls{FR3} \glspl{bs}. 
Differently from available models, we combine the power modeling of individual hardware elements 
in digital signal processing, analog signal processing, and \glspl{pa} with the averaging 
of each component's power during different phases (from data transmission/reception to idle mode) 
in a time frame. Moreover, we take into account hybrid analog-digital beamforming architectures.
The model is available in open access at this link 
\href{https://github.com/emanuele-peschiera}{\faGithubSquare}.\footnote{The code will 
be made available upon acceptance of the paper.}
Our main focus is on quantifying the impact of using large bandwidths (several hundreds of MHz), 
many \gls{bs} antenna elements (up to thousands) and RF chains. The link to the ergodic data rates 
in \gls{dl} and \gls{ul} sheds light on the relation between power consumption and capacity at \gls{FR3}.

\section{System Model}

\begin{figure*}[!t]
    \centering
    \includegraphics[trim={0 0em 0 0em},clip,width=.81\linewidth]{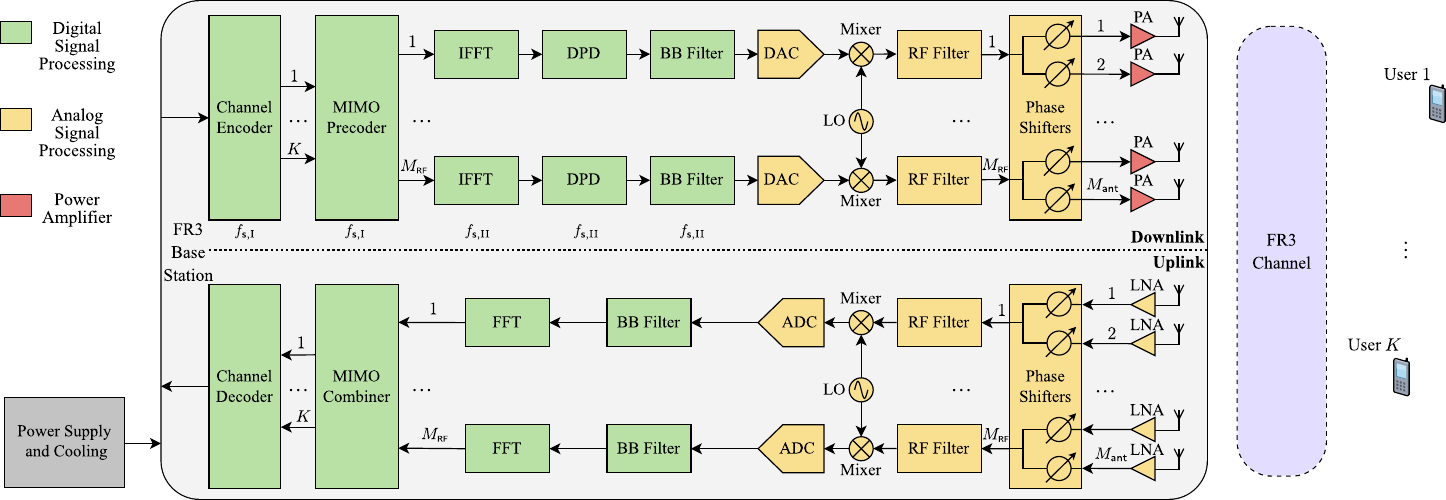}
    \vspace{-.2cm}
    \caption{Illustration of \gls{bs} architecture and system model considered in this work.}
    \vspace{-.2cm}
    \label{fig:BS}
\end{figure*}

As depicted in Fig.~\ref{fig:BS}, we consider a \gls{bs} equipped with $2\Mant$ antennas, \glspl{pa} and \glspl{lna}, 
and $2\MRF$ \gls{rf} chains that are equally split for use in \gls{dl} and \gls{ul}. 
The \gls{bs} serves $K$ single-antenna users using \gls{sdma} and \gls{ofdm}. 
The carrier frequency is $\sublow{f}{c}$ and the bandwidth is $B$.
We consider a frame of $\sublow{T}{f}$ seconds including \gls{dl} and \gls{ul}, where the time ratios of \gls{dl} and
\gls{ul} duration to $\sublow{T}{f}$ are given by $\subupp{\tau}{DL}\in[0,1]$ and $\subupp{\tau}{UL}\in[0,1]$.
As in~\cite{Golard_2024}, we define the instantaneous physical resource load at user $k$ and time 
$t\in\mathbb{R}$ as the ratio of the number of used data subcarriers ($\subupp{q}{DL}(k,t)$ in \gls{dl} and 
$\subupp{q}{UL}(k,t)$ in \gls{ul}) to available data subcarriers ($\subupp{Q}{DL}(k,t)$ in \gls{dl} and 
$\subupp{Q}{UL}(k,t)$ in \gls{ul}).
The average physical resource load in \gls{dl} and \gls{ul} is therefore given by
\vspace{-.15cm}
\begin{equation}\label{eq:load}
    \overline{x_i} = \frac{1}{\sublow{T}{f}}\int_0^{\sublow{T}{f}} \left(\frac{1}{K}
    \sum_{k=1}^K \frac{q_i(k,t)}{Q_i(k,t)}\right) \mathrm{d}t, 
    \quad i\in\{\mathsf{\scriptstyle{DL}},\mathsf{\scriptstyle{UL}}\}.
\end{equation}
\vspace{-.25cm}

\noindent By considering $i\in\{\mathsf{\scriptstyle{DL}},\mathsf{\scriptstyle{UL}}\}$, the \gls{dl} and \gls{ul} 
phases are further subdivided in: (i) data and reference signal transmission or reception, lasting 
$\overline{x_i}\tau_i\sublow{T}{f}$ seconds,
(ii) reference signal transmission or reception, lasting
$(1-\overline{x_i})\tau_i\sublow{\tau}{\mathit{i}\scriptstyle{,sig}}\sublow{T}{f}$ seconds, 
where $\sublow{\tau}{\mathit{i}\scriptstyle{,sig}}\in[0,1]$ is the time ratio of \gls{dl} or \gls{ul} signalling to 
\gls{dl} or \gls{ul} duration, and (iii) no data and no reference signal transmission or reception, lasting 
$(1-\overline{x_i})\tau_i(1-\sublow{\tau}{\mathit{i}\scriptstyle{,sig}})\sublow{T}{f}$ seconds and referred to as
micro-sleep mode. When not in \gls{dl} or \gls{ul} mode, the \gls{dl} 
or \gls{ul} \gls{bs} components are in idle mode. An illustration of the time frame in a similar scenario 
is given in~\cite[Fig.~1]{Peschiera_2025}.

Let us now describe the communication at one subcarrier and \gls{ofdm} symbol to ease notation, where
we consider that all the users are allocated $q_i$ subcarriers, 
$i\in\{\mathsf{\scriptstyle{DL}},\mathsf{\scriptstyle{UL}}\}$.
The subcarrier dependency will be considered in the sum rate expression.
The channel between the \gls{bs} and the user $k$
is denoted by $\vect{h}_k\inC{\Mant\times1}$, hence the channel matrix 
is given by $\mat{H}=\tp{[\vect{h}_1,\dotsc,\vect{h}_K]}\inC{K\times\Mant}$ and is assumed to be 
reciprocal during a channel coherence block. The users' symbols in \gls{dl} are
$\subupp{\vect{s}}{DL}\inC{K\times1}$ and satisfy 
$\expt{\subupp{\vect{s}}{DL}\hr{\subupp{\vect{s}}{DL}}}=(\subupp{P}{T,DL}/\subupp{q}{DL})\mat{I}_K$,
where is $\subupp{P}{T,DL}$ is the total transmit power at the \gls{bs}.
The \gls{bs} performs hybrid analog-digital precoding and transmits the signal $\sublow{\mat{W}}{ana}
\sublow{\mat{W}}{dig}\vect{s}\inC{\Mant\times1}$, where $\sublow{\mat{W}}{ana}\inC{\Mant\times\MRF}$ and
$\sublow{\mat{W}}{dig}=[\vect{w}_{\mathsf{dig},1},\dotsc,\vect{w}_{\mathsf{dig},K}]
\inC{\MRF\times K}$ are the analog and digital precoding matrices, respectively.
The analog precoding matrix contains phase-only elements and is selected from a codebook 
$\sublow{\mathcal{W}}{ana}$.
Furthermore, the precoding matrices are normalized such that 
$\norm{\sublow{\mat{W}}{ana}\sublow{\mat{W}}{dig}}_{\mathsf{\scriptscriptstyle{F}}}^2=1$.
The signal received by user $k$ is expressed as
$\tp{\vect{h}_k}\sublow{\mat{W}}{ana}\sublow{\mat{W}}{dig}\subupp{\vect{s}}{DL}+n_{\mathsf{\scriptscriptstyle{DL}},k}$,
where $n_{\mathsf{\scriptscriptstyle{DL}},k}$ is zero-mean circularly-symmetric complex Gaussian noise with variance 
$\sigma_{n,\mathsf{\scriptscriptstyle{DL}}}^2$.
Hence, the \gls{dl} \gls{sinr} of user $k$ is
\vspace{-.10cm}
\begin{equation}
    \!\!\subupp{\mathsf{SINR}}{DL\scriptstyle{,\mathit{k}}} = 
    \frac{\subupp{P}{T,DL}\abs{\tp{\vect{h}_k}\sublow{\mat{W}}{ana}\vect{w}_{\mathsf{dig},k}}^2}
    {\subupp{P}{T,DL}\sum_{k'=1,k'\neq k}^K \abs{\tp{\vect{h}_k}\sublow{\mat{W}}{ana}\vect{w}_{\mathsf{dig},k'}}^2+
    \subupp{q}{DL}\sigma_{n,\mathsf{\scriptscriptstyle{DL}}}^2}.
\end{equation}
\vspace{-.15cm}

In \gls{ul}, the users transmit $\subupp{\vect{s}}{UL}\inC{K\times1}$, with
$\expt{\subupp{\vect{s}}{UL}\hr{\subupp{\vect{s}}{UL}}}=(\subupp{P}{T,UL}/\subupp{q}{UL})\mat{I}_K$, 
$\subupp{P}{T,UL}$ being the transmit power of each user.
The \gls{bs} performs hybrid analog-digital combining via
the matrices $\sublow{\mat{V}}{dig}=\tp{[\vect{v}_{\mathsf{dig},1},\dotsc,\vect{v}_{\mathsf{dig},K}]}\inC{K\times\MRF}$ 
and $\sublow{\mat{V}}{ana}\inC{\MRF\times\Mant}$, where $\sublow{\mat{V}}{ana}$ is
selected from a codebook $\sublow{\mathcal{V}}{ana}$. The \gls{bs} obtains the symbol estimate of user $k$ as
$\tp{\vect{v}_{\mathsf{dig},k}}\sublow{\mat{V}}{ana}\hr{\mat{H}}\subupp{\vect{s}}{UL}+
\tilde{n}_{\mathsf{\scriptscriptstyle{UL}},k}$,
where $\tilde{n}_{\mathsf{\scriptscriptstyle{UL}},k}=\tp{\vect{v}_{\mathsf{dig},k}}\sublow{\mat{V}}{ana}\subupp{\vect{n}}{UL}$
with variance $\sigma_{\tilde{n},\mathsf{\scriptscriptstyle{UL}}}^2$, and
$\subupp{\vect{n}}{UL}$ is zero-mean circularly-symmetric complex-Gaussian noise with variance 
$\sigma_{n,\mathsf{\scriptscriptstyle{UL}}}^2$.
Hence, the \gls{ul} \gls{sinr} for user $k$ reads
\vspace{-.10cm}
\begin{equation}
    \subupp{\mathsf{SINR}}{UL\scriptstyle{,\mathit{k}}} = 
    \frac{\subupp{P}{T,UL}\abs{\tp{\vect{v}_{\mathsf{dig},k}}\sublow{\mat{V}}{ana}\vect{h}_k^*}^2}
    {\subupp{P}{T,UL}\sum_{k'=1,k'\neq k}^K \big|\tp{\vect{v}_{\mathsf{dig},k}}\sublow{\mat{V}}{ana}\vect{h}_{k'}^*\big|^2+
    \subupp{q}{UL}\sigma_{\tilde{n},\mathsf{\scriptscriptstyle{UL}}}^2}.
\end{equation}
\vspace{-.15cm}

\noindent The introduced SINRs depend, via the channel matrices and consequently the 
digital precoders/combiners, on the subcarrier index $\nu\in\{0,\dotsc,q_i\}$, therefore we
compute an achievable ergodic sum rate (in bit/s) in \gls{dl} and \gls{ul} with effective bandwidth $\tilde{B}$,
as
\vspace{-.10cm}
\begin{equation}
    \!\!\!\!R_i = \overline{x_i}\tau_i
    \frac{\tilde{B}}{q_i}\,\mathbb{E}\bigg\{\sum_{\nu=1}^{q_i}\sum_{k=1}^K\log_2\Big(1+\mathsf{SINR}_{i,k}[\nu]\Big)\bigg\},
    \ i\!\in\!\{\mathsf{\scriptstyle{DL}},\mathsf{\scriptstyle{UL}}\}.
\end{equation}
\vspace{-.35cm}

\section{Power Model for a FR3 Base Station}
In this section, we describe the main \gls{bs} components and subcomponents, and their power consumption models.
We subdivide the \gls{bs} components in digital processing, analog processing, and \glspl{pa}, 
and express the total \gls{bs} power consumption averaged over the frame as
\vspace{-.15cm}
\begin{equation}
    \Pcons = \frac{\overline{\sublow{P}{dig}}}{\sublow{\eta}{dig,s/c}}+
    \frac{\overline{\sublow{P}{ana}}}{\sublow{\eta}{ana,s/c}}+
    \frac{\overline{\subupp{P}{PA}}}{\subupp{\eta}{PA\scriptstyle{,s/c}}}
\end{equation}
\vspace{-.25cm}

\noindent where $\sublow{\eta}{dig,s/c}\in(0,1]$, $\sublow{\eta}{ana,s/c}\in(0,1]$, and
$\subupp{\eta}{PA\scriptstyle{,s/c}}\in(0,1]$ are the supply and cooling efficiencies of each component.

\subsection{Digital Signal Processing}

The digital signal processing part of a \gls{bs} receives the information bits and feeds the 
\glspl{dac} in \gls{dl}, while it receives the \glspl{adc} output and returns the information bits in 
\gls{ul}. The digital subcomponents that we consider in this paper are illustrated in green in Fig.~\ref{fig:BS}.
The average digital consumption in both \gls{dl} and \gls{ul} is quantified as
$\overline{\sublow{P}{dig}} = 
\overline{\sublow{P}{dig,\scriptscriptstyle{DL}}}+
\overline{\sublow{P}{dig,\scriptscriptstyle{UL}}}$,
with
\vspace{-.15cm}
\begin{equation}\label{eq:P_idig}
\begin{aligned}
    \overline{\sublow{P}{dig,\mathit{i}}} =&\; \overline{\subupp{x}{\mathit{i}}}\tau_i\sublow{P}{dig,\mathit{i}}
    +\tau_i\sublow{\tau}{\mathit{i},sig}\sublow{P}{dig,\mathit{i}}\\
    +&\; \tau_i(1-\overline{x_i})(1-\sublow{\tau}{\mathit{i},sig})
    \sublow{P}{dig,\mathit{i}}\sublow{\delta}{dig,micro}\\
    +&\; (1-\tau_i)\sublow{P}{dig,\mathit{i}}\sublow{\delta}{dig,idle},\quad i\in\{\mathsf{\scriptstyle{DL}},\mathsf{\scriptstyle{UL}}\}
\end{aligned}
\end{equation}
\vspace{-.25cm}

\noindent being the digital consumption in DL or UL averaged over the frame, 
where $\sublow{\delta}{dig,micro}\in[0,1]$ and $\sublow{\delta}{dig,idle}\in[0,1]$ are reduction factors when 
the digital subcomponents are in micro-sleep and idle mode, respectively. 
The frame-averaged consumption is therefore computed as a weighted sum of the consumption during each phase of the frame.
The consumption when the digital processing is active is
$\sublow{P}{dig,\scriptscriptstyle{DL}} = 
\sublow{P}{encoder}+\sublow{P}{precoder}+\subupp{P}{IFFT}+
\subupp{P}{DPD}+\sublow{P}{filter,\scriptscriptstyle{BB}}$
in DL, and
$\sublow{P}{dig,\scriptscriptstyle{UL}} = 
\sublow{P}{decoder}+\sublow{P}{combiner}+\subupp{P}{FFT}+
\sublow{P}{filter,\scriptscriptstyle{BB}}$ in UL.
We propose to set in the evaluations $\sublow{\delta}{dig,micro}=0.5$ and $\sublow{\delta}{dig,idle}=0.25$.

To quantify the power consumption of a subcomponent, we include a static 
(signal-independent) and a dynamic (signal-dependent) term, which are
influenced by the hardware architectural choice. The static term 
is indicated by the subscript `$\mathsf{s}$', while the dynamic term is 
computed as the number of complex \gls{gops} (indicated by $\sublow{\Xi}{component}$) divided by
the computational efficiency (indicated by $\sublow{\eta}{component}$) in complex \gls{gops}/W.
The channel encoding/decoding and the \gls{mimo} precoding/combining
operate at the constellation rate, denoted as $\sublow{f}{s,\mathrm{I}}$, while the
other subcomponents operate at a sampling rate larger than $\sublow{f}{s,\mathrm{I}}$, 
denoted as $\sublow{f}{s,\mathrm{II}}$.\footnote{
These are equal to $\sublow{f}{s,\mathrm{I}}=\mu B$, $\mu\in[0,1)$ and we consider $\mu=0.9$, and
$\sublow{f}{s,\mathrm{II}}=\subupp{Q}{IFFT}\Delta f$ where $\subupp{Q}{IFFT}$
is the \gls{ifft} size and $\Delta f$ is the subcarrier spacing, 
set to $\subupp{Q}{IFFT}=4096$, $\Delta f=120$ kHz.}

The most common hardware solutions for cellular wireless are \gls{fpga} and \gls{asic}.
To estimate the computational efficiency and static consumption of \gls{fpga}
solutions, we implemented a \gls{mimo} precoder based on LDL decomposition on an 
AMD Zynq UltraScale+ \gls{rfsoc} ZCU216~\cite{zcu216}. With $32$ antennas, $8$ users,
and clock frequency of $200$ MHz, the computations of precoding matrices and precoded signals are performed at
$0.8$ MHz and $400$ MHz, respectively, where we estimate $\eta=0.2\cdot10^{3}$ complex GOPS/W and a static consumption of $1$ W.
For \gls{asic} implementations, we consider a ten times larger $\eta$ and a ten times lower 
static consumption~\cite{Boutros_2021}.

Let us now express the power consumption of each subcomponent when in active mode, i.e., transmitting or
receiving data or reference signals. 
Using the values in~\cite{Desset_2020} for \gls{ldpc} codes, we can write for the channel encoder and decoder
\vspace{-.15cm}
\begin{equation}
    \sublow{P}{\textit{i}} =
    \sublow{P}{\textit{i},s}+
    \frac{\sublow{f}{s,\mathrm{I}}}{\sublow{\eta}{\textit{i}}}
    \sublow{\Xi}{\textit{i}},\quad i\in\{\mathsf{\scriptstyle{encoder}},\mathsf{\scriptstyle{decoder}}\} 
\end{equation}
\vspace{-.25cm}

\noindent where $\sublow{\Xi}{encoder}=\frac{14}{3\cdot8}\big(\subupp{R}{DL}/\tilde{B}\big)$ and
$\sublow{\Xi}{decoder}=\frac{5\cdot35}{2\cdot3}\big(\subupp{R}{UL}/\tilde{B}\big)$.
We consider \gls{asic} implementation giving $\sublow{P}{encoder,s}=\sublow{P}{decoder,s}=0.1$ W
and $\sublow{\eta}{encoder}=\sublow{\eta}{decoder}=0.2\cdot10^{4}$ GOPS/W.
For the \gls{mimo} precoding and combining, we 
assume that a \gls{zf} forcing scheme is used.
Given the channel reciprocity, we have
$\sublow{\mat{W}}{ana} = \hr{\sublow{\mat{V}}{ana}}$, hence
$\sublow{\mat{W}}{dig} = \hr{\sublow{\mat{V}}{dig}} = \hr{\sublow{\mat{H}}{eff}}
(\sublow{\mat{H}}{eff}\hr{\sublow{\mat{H}}{eff}})^{-1}$,
where $\sublow{\mat{H}}{eff} = \mat{H}\sublow{\mat{W}}{ana}$.
Due to the required flexibility to change the precoding/combining scheme
or adapt it to different scenarios, we select
\gls{fpga} as hardware architecture giving $\sublow{P}{precoder,s}=\sublow{P}{combiner,s}=1$ W 
and $\sublow{\eta}{precoder}=\sublow{\eta}{combiner}=0.2\cdot10^{3}$ GOPS/W.
By considering the computation of precoded signals and precoding matrices, 
where a single \gls{fpga} is shared among $32$ antennas, we can write
\vspace{-.15cm}
\begin{equation}
\begin{aligned}
    \sublow{P}{\mathit{i}} &=
    \subupp{\tilde{M}}{RF}\sublow{P}{\mathit{i},s}+
    \frac{\sublow{f}{s,\mathrm{I}}}{\sublow{\eta}{\mathit{i}}}
    \MRF\sublow{\Xi}{\mathit{i}},\quad i\in\{\mathsf{\scriptstyle{precoder}},\mathsf{\scriptstyle{combiner}}\} 
\end{aligned}
\end{equation}
\vspace{-.25cm}

\noindent where $\subupp{\tilde{M}}{RF}\!=\!\big\lceil\frac{\MRF}{32}\big\rceil$,
$\sublow{\Xi}{precoder}=2K+\sublow{\upsilon}{coh}^{-1}\big(K^3/(3\MRF)+
3K^2+ K\big)$, and $\sublow{\Xi}{combiner}=2K$,
$\sublow{\upsilon}{coh}$ being the number of samples per coherence block.
For \gls{ofdm} modulation/demodulation, we assume 5G-like operation
where the number of subcarriers is fixed and different subcarrier spacings give
different bandwidths~\cite{dahlman20205g}, and consider that they are implemented in \gls{asic}. 
We then express
\vspace{-.15cm}
\begin{equation}
    \subupp{P}{IFFT} = \subupp{P}{FFT} = 
    \subupp{P}{IFFT\scriptstyle{,s}}+
    \frac{\sublow{f}{s,\mathrm{II}}}{\subupp{\eta}{IFFT}}
    \MRF\subupp{\Xi}{IFFT}
\end{equation}
\vspace{-.25cm}

\noindent where $\subupp{\Xi}{IFFT}=\frac{3}{2}\log_2\subupp{Q}{IFFT}$ and $\subupp{Q}{IFFT}$ is the \gls{ifft}/FFT size
that we fix to $\subupp{Q}{IFFT}=4096$. 
Next, the \gls{dpd} is tipically characterized by large \gls{gops},
both for filtering and coefficients computation. Its power consumption is modeled as
\vspace{-.15cm}
\begin{equation}
    \subupp{P}{DPD} = 
    \subupp{P}{DPD\scriptstyle{,s}}+
    \frac{\sublow{f}{s,\mathrm{II}}}{\subupp{\eta}{DPD}}
    \MRF\subupp{\Xi}{DPD}
\end{equation}
\vspace{-.25cm}

\noindent where we use~\cite{Cao_2022} to set $\subupp{\Xi}{DPD}=50$
and consider it is implemented in \gls{asic}.
Last, the \gls{bb} filter performs interpolation in \gls{dl}
and decimation in \gls{ul}. Given that \gls{FR3} subbands are
expected to have different spectral requirements, we opt for 
the \gls{fpga} solution~\cite{Mezzavilla_2024}. 
We model its power consumption as
\vspace{-.15cm}
\begin{equation}
    \sublow{P}{filter,\scriptscriptstyle{BB}} = 
    \subupp{\tilde{M}}{RF}\sublow{P}{filter,\scriptscriptstyle{BB}\scriptstyle{,s}}+
    \frac{\sublow{f}{s,\mathrm{II}}}{\sublow{\eta}{filter,\scriptscriptstyle{BB}}}
    \MRF\sublow{\Xi}{filter,\scriptscriptstyle{BB}}
\end{equation}
\vspace{-.25cm}

\noindent where $\sublow{\Xi}{filter,\scriptscriptstyle{BB}}=\sublow{n}{filter,\scriptscriptstyle{BB}}
\sublow{o}{filter,\scriptscriptstyle{BB}}$,
with $\sublow{n}{filter,\scriptscriptstyle{BB}}$ and
$\sublow{o}{filter,\scriptscriptstyle{BB}}$ being the number of filter taps 
and oversampling coefficient, which we set to $\sublow{n}{filter,\scriptscriptstyle{BB}}=20$
and $\sublow{o}{filter,\scriptscriptstyle{BB}}=4$.

\subsection{Analog Signal Processing}
In the analog signal processing, we consider the subcomponents illustrated in orange in Fig.~\ref{fig:BS}.
The \gls{lo} is shared among the \gls{rf} chains and the \gls{lna} is included in the 
\gls{ul} subcomponents. The focus is on hybrid partially-connected beamforming architectures 
with $\subupp{M}{PS}=\Mant/\MRF$ phase shifters per \gls{rf} chain 
(equal to the number of antennas per subarray) and fully-digital beamforming architectures with $\subupp{M}{PS}=0$.
Similarly to the digital consumption, the average analog consumption in both \gls{dl} and \gls{ul} is computed via
$\overline{\sublow{P}{ana}} = 
\overline{\sublow{P}{ana,\scriptscriptstyle{DL}}}+
\overline{\sublow{P}{ana,\scriptscriptstyle{UL}}}$, where
\vspace{-.15cm}
\begin{equation}\label{eq:P_iana}
\begin{aligned}
    \overline{\sublow{P}{ana,\mathit{i}}} =&\; \overline{x_i}\tau_i\sublow{P}{ana,\mathit{i}}
    + \tau_i\sublow{\tau}{\mathit{i}\scriptstyle{,sig}}\sublow{P}{ana,\mathit{i}}\\
    +&\; \tau_i(1-\overline{x_i})(1-\sublow{\tau}{\mathit{i},sig})
    \sublow{P}{ana,\mathit{i}}\sublow{\delta}{ana,micro}\\
    +&\; (1-\tau_i)\sublow{P}{ana,\mathit{i}}\sublow{\delta}{ana,idle},
    \quad i\in\{\mathsf{\scriptstyle{DL}},\mathsf{\scriptstyle{UL}}\}
\end{aligned}
\end{equation}
\vspace{-.25cm}

\noindent is the the analog processing consumption in \gls{dl} or \gls{ul} averaged over the frame, and
$\sublow{\delta}{ana,micro}\in[0,1]$ and $\sublow{\delta}{ana,idle}\in[0,1]$ are reduction factors when
the analog subcomponents enter micro-sleep and idle modes, respectively.
The total analog processing consumption during active mode in \gls{dl} and \gls{ul} is given by
$\sublow{P}{ana,\scriptscriptstyle{DL}} = \subupp{P}{LO}+\MRF\sublow{P}{ana,\scriptscriptstyle{DL}\scriptstyle{,1}}$
and $\sublow{P}{ana,\scriptscriptstyle{UL}} = \subupp{P}{LO}+\MRF\sublow{P}{ana,\scriptscriptstyle{UL}\scriptstyle{,1}}$,
where $\sublow{P}{ana,\scriptscriptstyle{DL}\scriptstyle{,1}} = 
2\subupp{P}{DAC}+2\sublow{P}{filter,\scriptscriptstyle{RF}}+
2\sublow{P}{mix}+\subupp{M}{PS}\subupp{P}{PS}$ and 
$\sublow{P}{ana,\scriptscriptstyle{UL}\scriptstyle{,1}} = 
2\subupp{P}{ADC}+2\sublow{P}{filter,\scriptscriptstyle{RF}}+
2\sublow{P}{mix}+\subupp{M}{PS}\subupp{P}{PS}+\subupp{P}{LNA}$ are the
power consumptions of one \gls{rf} chain in \gls{dl} and \gls{ul}.
In the previous expressions, we include two (I and Q channels) \glspl{dac}, \glspl{adc}, 
\gls{rf} filters, and mixers.
In the following, we propose to set $\sublow{\delta}{ana,micro}=0.75$ and $\sublow{\delta}{ana,idle}=0.5$.

We consider that the \gls{dac} and \gls{adc} operate at large (\gls{rf}) sampling frequencies to enable
full control in the digital processing part.
A model for the \gls{dac} power consumption is given by~\cite{Shuguang_2005}
\vspace{-.15cm}
\begin{equation}
\subupp{P}{DAC}  = \subupp{P}{DAC\scriptstyle{,s}}+1.5\cdot10^{-12}\subupp{b}{DAC}\subupp{f}{DAC}
\end{equation}
\vspace{-.45cm}

\noindent where $\subupp{P}{DAC\scriptstyle{,s}}=1.5\cdot10^{-5}2^{\subupp{b}{DAC}}$, $\subupp{b}{DAC}$ is the effective number of bits 
and $\subupp{f}{DAC}$ is the sampling frequency. This models well state-of-the-art high-speed \glspl{dac} such as~\cite{Sihao_2025},
and we set in the following $\subupp{b}{DAC}=8$ and $\subupp{f}{DAC}=5$ GS/s.
The \gls{adc} power consumption can be generally modeled as~\cite{adc_survey}
\vspace{-.15cm}
\begin{equation}
\subupp{P}{ADC} = \subupp{\mathsf{W}}{ADC}\subupp{f}{ADC}2^{\subupp{b}{ADC}}
\end{equation}
\vspace{-.25cm}

\noindent where $\subupp{\mathsf{W}}{ADC}$ is the Walden's figure-of-merit, ${\subupp{b}{ADC}}$ is the effective number of bits 
and $\subupp{f}{ADC}$ is the sampling frequency. State-of-the-art high-speed \glspl{dac} have $\subupp{\mathsf{W}}{ADC}$
around $70$ fJ/cs~\cite{Markulic_2024}, where we consider $\subupp{b}{ADC}=8$ and $\subupp{f}{ADC}=5$ GS/s.
We express the consumption of a mixer, phase shifter, and \gls{lna} as~\cite{Helal_2025},\cite{Abbasi_2025},\cite{Afifi_2025}
\vspace{-.15cm}
\begin{equation}
\sublow{P}{mix} = \sublow{\Xi}{mix}\sublow{f}{c},\ \subupp{P}{PS} = \subupp{\Xi}{PS}B,\ \subupp{P}{LNA} = \subupp{\Xi}{LNA}B
\end{equation}
\vspace{-.35cm}

\noindent where $\sublow{\Xi}{mix}=2.5\cdot10^{-13}$, $\subupp{\Xi}{PS}=3.5\cdot10^{-11}$ is a figure-of-merit that depends on gain, 
noise factor, and resolution, and $\subupp{\Xi}{LNA}=2.7\cdot10^{-11}$ is a figure-of-merit that considers gain and noise factor.
For the consumption of the \gls{lo} and the \gls{rf} filter,
we use the datasheets in~\cite{ADMV8818},~\cite{ADF41510} and fix them to 
$\subupp{P}{LO}=40$ mW and $\sublow{P}{filter,\scriptscriptstyle{RF}}=5$ mW.

\begin{figure*}[!t]
    \centering
    \footnotesize
    In \textbf{(a)} and \textbf{(c)}: $\Mant=1024$, $B=400$ MHz, 
    $\subupp{P}{T,DL}=100$ W, $\subupp{P}{T,UL}=100$ mW, $K=8$.\ \
    In \textbf{(a)}, \textbf{(b)}, and \textbf{(c)}:
    $\subupp{\tau}{DL}=0.75$, $\subupp{\tau}{UL}=0.25$ 
    \begin{minipage}{.35\textwidth}
    \pgfplotsset{grid style={dotted,gray}}

\begin{tikzpicture}[
    every axis/.style={
      ybar stacked,
      ymin=0, ymax=750,
      bar width=6pt,
      width=5cm,
      height=5cm,
      ylabel={Power consumption, $\Pcons$ [W]},
      ylabel style={yshift=-.5ex},
      ytick={0,100,200,300,400,500,600,700,800,1000},
      yticklabels={$0$,$100$,$200$,$300$,$400$,$500$,$600$,$700$,$800$,$1000$},
      xlabel={Number of RF chains, $\MRF$},
      xtick={1,2,3,4,5,6,7},
      xlabel style={yshift=.5ex},
      xticklabels={$16$${^\star}$,$32$${^\star}$,$64$${^\star}$,$128$${^\star}$,$256$${^\star}$,$512$${^\star}$,$1024$$^\circ$},
    },
]

\footnotesize

\definecolor{mygreen}{RGB}{166,216,159}
\definecolor{myorange}{RGB}{255,213,128}
\definecolor{myred}{RGB}{204,76,67}

\begin{axis}[
    width=6.5cm,
    height=5cm,
    grid=both,
    bar shift=-3pt
]

\addplot[fill=mygreen!150] 
coordinates {(1,5.32) (2,8.31) (3,16.22) (4,32.05) (5,63.71) (6,127.02) (7,253.64)};
\addplot[fill=mygreen!75] 
coordinates {(1,3.62) (2,5.67) (3,11.1) (4,21.96) (5,43.68) (6,87.12) (7,173.99)};
\end{axis}

\begin{axis}[
    width=6.5cm,
    height=5cm,
    hide axis,
    bar shift=3pt,
    legend columns=1,
    legend style={nodes={scale=0.8},at={(0,1)},anchor=north west},
]
\addplot[fill=myorange!150,postaction={pattern=dots,pattern color=white},forget plot] 
coordinates {(1,29.56) (2,36.62) (3,50.73) (4,78.96) (5,135.41) (6,248.32) (7,451.7)};
\addplot[fill=myorange!75,postaction={pattern=dots,pattern color=white}, forget plot] 
coordinates {(1,6.7) (2,8.01) (3,10.63) (4,15.86) (5,26.34) (6,47.29) (7,83.82)};

\addlegendimage{fill=mygreen!150}
\addlegendentry{Digital (load-indep.)}
\addlegendimage{fill=mygreen!75}
\addlegendentry{Digital (load-dep.)}
\addlegendimage{fill=myorange!150,postaction={pattern=dots,pattern color=white}}
\addlegendentry{Analog (load-indep.)}
\addlegendimage{fill=myorange!75,,postaction={pattern=dots,pattern color=white}}
\addlegendentry{Analog (load-dep.)}

\draw (axis cs:5.4,210) node[anchor=east,rotate=90.0,right,align=center,font=\linespread{0.4}\selectfont] 
{\tiny $\overline{\subupp{x}{\scriptscriptstyle{DL}}}+\overline{\subupp{x}{\scriptscriptstyle{UL}}}\in[0,2]$};
\draw[latex'-latex'] (axis cs:5.55,127.02) -- (axis cs:5.55,127.02+87.12);

\end{axis}

\end{tikzpicture}
    \end{minipage}
    \hspace{-.35cm}
    \begin{minipage}{.25\textwidth}\vspace{.125cm}
    \pgfplotsset{grid style={dotted,gray}}
\newcommand{\marker}{\raisebox{0.5pt}{\tikz{\node[draw,scale=0.45,diamond,fill=white](){};}}}

\begin{tikzpicture}[
    every axis/.style={
      ymin=0, ymax=750,
      bar width=6pt,
      width=5.5cm,
      height=5cm,
      grid=both,
      ylabel={},
      ylabel style={yshift=-.5ex},
    },
]

\footnotesize

\definecolor{mygreen}{RGB}{166,216,159}
\definecolor{myorange}{RGB}{255,213,128}
\definecolor{myred}{RGB}{204,76,67}

\begin{axis}[
    ybar stacked,
    width=5.5cm,
    height=5cm,
    bar shift=0pt,
    grid=both,
    ymin=0, ymax=750,
    legend pos=north west,
    legend columns=1,
    legend style={nodes={scale=0.8},at={(0,1)},anchor=north west},
    ytick={0,100,200,300,400,500,600,700,800,1000},
    yticklabels={},
    xlabel={Number of antennas, $\Mant$},
    xtick={1,2,3,4,5,6,7},
    xlabel style={yshift=.5ex},
    xticklabels={$16$,$32$,$64$,$128$,$256$,$512$,$1024$}
]
\addplot[fill=myred!150,forget plot,postaction={pattern=north east lines,pattern color=white},forget plot] 
coordinates {(1,1.19) (2,2.37) (3,4.74) (4,9.49) (5,18.97) (6,37.94) (7,75.89)};
\addplot[fill=myred!75,forget plot,postaction={pattern=north east lines,pattern color=white},forget plot] 
coordinates {(1,9.68) (2,19.35) (3,38.7) (4,77.41) (5,154.82) (6,309.63) (7,619.08)};

\addlegendimage{fill=myred!150,postaction={pattern=north east lines,pattern color=white}}
\addlegendentry{Power amplifier (load-indep.)}
\addlegendimage{fill=myred!75,postaction={pattern=north east lines,pattern color=white}}
\addlegendentry{Power amplifier (load-dep.)}

\draw[latex'-latex'] (axis cs:5.65,37.94) -- (axis cs:5.65,309.63+37.94);

\draw (axis cs:0.6,520) node[anchor=east,rotate=0,right,align=center,font=\linespread{0.4}\selectfont] 
{\tiny \marker\ \ is at load $\overline{\subupp{x}{DL}}=0.3$};

\end{axis}

\begin{axis}[
    hide axis,
    width=5.5cm,
    height=5cm,
    bar shift=0pt,
    ymin=0, ymax=750,
    xmin=0.4, xmax=7.6,
    legend pos=north west,
    legend columns=1,
    legend style={nodes={scale=0.8}}
]
\addplot[only marks,mark=diamond*,mark size=2,mark options={fill=white,draw=black},forget plot,
] coordinates {
  (4,0.3*77.41+9.49)
  (5,0.3*154.82+18.97)
  (6,0.3*309.63+37.94)
  (7,0.3*619.08+75.89)
};
\end{axis}

\end{tikzpicture}
    \end{minipage}
    \begin{minipage}{.35\textwidth}\vspace{.1cm}
    \pgfplotsset{grid style={dotted,gray}}
\newcommand{\markerone}{\raisebox{0.5pt}{\tikz{\node[draw,scale=0.52,regular polygon, regular polygon sides=4,fill=white](){};}}}
\newcommand{\markertwo}{\raisebox{0.5pt}{\tikz{\node[draw,scale=0.35,regular polygon, regular polygon sides=3,fill=white](){};}}}

\begin{tikzpicture}[
    every axis/.style={
      ymin=0,
      bar width=6pt,
      width=5cm,
      height=5cm,
      ylabel={Ergodic sum rate, $R_i$ [Gbit/s]},
      ylabel style={yshift=-.5ex},
      xlabel={Number of RF chains, $\MRF$},
      xlabel style={yshift=.5ex},
      ytick={0,2,4,6,8,10,12},
      yticklabels={$0$,$2$,$4$,$6$,$8$,$10$,$12$},
    },
]

\footnotesize

\definecolor{myblue1}{RGB}{70,130,180}
\definecolor{myblue2}{RGB}{135,206,235}

\begin{axis}[
      width=6.5cm,
      height=5cm,
      ybar stacked,
      xtick={1,2,3,4,5,6,7},
      xticklabels={$16$${^\star}$,$32$${^\star}$,$64$${^\star}$,$128$${^\star}$,$256$${^\star}$,$512$${^\star}$,$1024$$^\circ$},
      grid=both,
      bar shift=-3pt,
      ymin=0, ymax=14
]
\addplot[fill=myblue2] 
coordinates {(1,1.02) (2,1.25) (3,1.5) (4,1.68) (5,1.98) (6,2.18) (7,2.35)};

\end{axis}

\begin{axis}[
    width=6.5cm,
    height=5cm,
    ybar stacked,
    bar shift=3pt,
    hide axis,
    ymin=0, ymax=14,
    legend pos=north west,
    legend columns=1,
    legend style={nodes={scale=0.8},at={(0,1)},anchor=north west}
]
\addplot[fill=myblue1,forget plot,postaction={pattern=north west lines,pattern color=white}] 
coordinates {(1,1.61) (2,3.64) (3,6.06) (4,8.10) (5,10.93) (6,12.10) (7,13.44)};

\addlegendimage{fill=myblue2}
\addlegendentry{Uplink, $i=\mathsf{UL}$}
\addlegendimage{fill=myblue1,postaction={pattern=north west lines,pattern color=white}}
\addlegendentry{Downlink, $i=\mathsf{DL}$}

\draw (axis cs:0.61,9.8) node[anchor=east,rotate=0,right,align=center,font=\linespread{0.4}\selectfont] 
{\tiny \markerone\ \ is at load $\overline{\subupp{x}{UL}}=0.3$};
\draw (axis cs:0.6,8.8) node[anchor=east,rotate=0,right,align=center,font=\linespread{0.4}\selectfont] 
{\tiny \markertwo\ \ is at load $\overline{\subupp{x}{DL}}=0.3$};

\end{axis}

\begin{axis}[
    hide axis,
    width=6.5cm,
    height=5cm,
    ymin=0, ymax=14,
    xmin=0.55, xmax=7.75
]
\addplot[only marks,mark=square*,mark size=1.4,mark options={fill=white,draw=black},forget plot,
] coordinates {
  (1,0.3*1.02)
  (2,0.3*1.25)
  (3,0.3*1.5)
  (4,0.3*1.68)
  (5,0.3*1.98)
  (6,0.3*2.18)
  (7,0.3*2.35)
};
\end{axis}

\begin{axis}[
    hide axis,
    width=6.5cm,
    height=5cm,
    ymin=0, ymax=14,
    xmin=0.25, xmax=7.45
]
\addplot[only marks,mark=triangle*,mark size=2,mark options={fill=white,draw=black},forget plot,
] coordinates {
  (1,0.3*1.61)
  (2,0.3*3.64)
  (3,0.3*6.06)
  (4,0.3*8.10)
  (5,0.3*10.93)
  (6,0.3*12.10)
  (7,0.3*13.44)
};

\end{axis}

\end{tikzpicture}
    \end{minipage}
    \\\hspace{0cm}\textbf{(a)}\hspace{4.5cm}\textbf{(b)}\hspace{5.2cm}\textbf{(c)}\\
    \vspace{-.3cm}
    \flushleft
    $^\star$Hybrid analog-digital beamforming with $\MRF\in\{16,32,\dotsc,512\}$ and $\Mant=1024$ \quad $^\circ$Fully-digital beamforming with $\MRF=\Mant=1024$\\
    \vspace{-.25cm}
    \caption{Evaluations of power consumption in \textbf{(a)} and \textbf{(b)} and ergodic rates in \textbf{(c)} for the considered \gls{FR3} scenario.}
    \vspace{-.2cm}
    \label{fig:eval}
\end{figure*}

\subsection{Power Amplifier}

Let us denote the average per-\gls{pa} (and per-antenna) transmit power by $\Pa=\subupp{P}{T,DL}/\Mant$.
By considering the frame structure as for the other components, the average consumption over the frame of the
$\Mant$ \glspl{pa} is given by $\overline{\subupp{P}{PA}} = \Mant\overline{\subupp{P}{PA\scriptstyle{,1}}}$,
with the average consumption of one \gls{pa} being expressed as
\vspace{-.15cm}
\begin{equation}\label{eq:P_pa}
\begin{aligned}
    \overline{\subupp{P}{PA\scriptstyle{,1}}} =&\;\overline{\subupp{x}{DL}}\subupp{\tau}{DL}\subupp{P}{PA\scriptstyle{,1}}(\Pa)
    +\subupp{\tau}{DL}\subupp{\tau}{DL\scriptstyle{,sig}}\subupp{P}{PA\scriptstyle{,1}}(\subupp{\zeta}{DL\scriptstyle{,sig}}\Pmax)\\
    +&\;\subupp{\tau}{DL}(1-\overline{\subupp{x}{DL}})(1-\subupp{\tau}{DL\scriptstyle{,sig}})\subupp{P}{PA\scriptstyle{,1}}(0)\subupp{\delta}{PA\scriptstyle{,micro}}\\
    +&\;(1-\subupp{\tau}{DL})\subupp{P}{PA\scriptstyle{,1}}(0)\subupp{\delta}{PA\scriptstyle{,idle}}
\end{aligned}
\end{equation}
\vspace{-.25cm}

\noindent where $\subupp{\zeta}{DL\scriptstyle{,sig}}\in(0,1)$ is the fraction of maximal output power that is used for signalling,
and $\subupp{\delta}{PA\scriptstyle{,micro}}\in[0,1]$ and $\subupp{\delta}{PA\scriptstyle{,idle}}\in[0,1]$ quantify the
power savings when the \gls{pa} is in micro-sleep and idle mode. These are set to $\subupp{\delta}{PA\scriptstyle{,micro}}=0.5$ 
and $\subupp{\delta}{PA\scriptstyle{,idle}}=0.25$~\cite{Cheng_2014}.
To model the instantaneous \gls{pa} consumption, we adopt the model in~\cite{Golard_2024}
\vspace{-.15cm}
\begin{equation}
    \subupp{P}{PA\scriptstyle{,1}}(p) = \xi\frac{\Pmax^\alpha}{\subupp{\eta}{PA\scriptstyle{,max}}}+
	(1-\xi)\frac{\Pmax^{1-\alpha}p^\alpha}{\subupp{\eta}{PA\scriptstyle{,max}}}
\end{equation}
\vspace{-.25cm}

\noindent where $p$ is the instantaneous output power, $\Pmax$ is the maximal \gls{pa} output power considering
a certain output power back-off (that we set to $8$ dB) from the \gls{pa} saturation power, $\xi\in[0,1]$ is the ratio of static 
to dynamic power consumption, $\subupp{\eta}{PA\scriptstyle{,max}}$ is the maximum \gls{pa} efficiency, 
and $\alpha\in[0,1]$. The above model considers a non-constant \gls{pa} efficiency that depends on $p$,
and its parameters can be adapted to a specific \gls{pa} technology.
Based on~\cite{Wang_2024}, the three main technologies for implementing \glspl{pa} at \gls{FR3} are
\gls{gan}, \gls{gaas}, \gls{sige}, and \gls{cmos}, giving different $\Pmax$ and $\subupp{\eta}{PA\scriptstyle{,max}}$ 
among other factors. Based on a recent \gls{sige} implementation at \gls{FR3}~\cite{Pecile_2025}, 
we set $\Pmax=0.1$ W, $\subupp{\eta}{PA\scriptstyle{,max}}=0.15$, $\alpha=0.75$, and $\xi=0.1$.

\vspace{-.4cm}
\section{Numerical Evaluation}
In this section, we present the evaluations of power consumption and ergodic rates in a \gls{FR3} scenario.
We consider a \gls{3gpp} 39.801 Urban Micro (UMi) wideband channel model generated by the Sionna simulator~\cite{sionna}. 
The \gls{bs} array is a uniform planar array (UPA) having up to $\Mant=1024$ antenna elements 
and half-wavelength spacing in both directions, and the codebook for analog beamforming 
$\sublow{\mathcal{W}}{ana}=\sublow{\mathcal{V}}{ana}$ is generated via a 2D-DFT. We set the transmit powers
at the \gls{bs} and at the user as $\subupp{P}{T,DL}=100\frac{\Mant}{1024}$ W such that
$\Pa\approx0.1=\Pmax$ W and $\subupp{P}{T,DL}=100$ W when there are $1024$ antennas, 
and $\subupp{P}{T,UL}=100$ mW. 
The carrier frequency, bandwidth, and effective bandwidth are 
$\sublow{f}{c}=10$ GHz, $B=400$ MHz, and $\tilde{B}=0.9 B$, while the number of data subcarriers is 
$\subupp{q}{DL}=\subupp{q}{UL}=3000$.
The number of users is $K=8$. The noise powers are set as 
$\subupp{\sigma}{\scriptstyle{\mathit{n},}DL}^2=\subupp{\sigma}{\scriptstyle{\mathit{n},}UL}^2=-123$ dBm.
We consider \gls{tdd} operation with $\subupp{\tau}{DL}=0.75$, $\subupp{\tau}{UL}=0.25$, 
$\subupp{\tau}{DL\scriptstyle{,sig}}=\subupp{\tau}{UL\scriptstyle{,sig}}=1/14$ and $\subupp{\zeta}{DL\scriptstyle{,sig}}=1/12$~\cite{Golard_2024}.
We futher set $\sublow{\eta}{dig,s/c}=\sublow{\eta}{ana,s/c}=\subupp{\eta}{PA\scriptstyle{,s/c}}=0.8$.
According to~(\ref{eq:load}), the average physical loads are bounded as 
$\overline{\subupp{x}{DL}}\in[0,1]$ and $\overline{\subupp{x}{UL}}\in[0,1]$.

In Fig.~\ref{fig:eval}, we show the power consumption and ergodic rates while varying either the 
number of RF chains or the number of antennas from $16$ to $1024$. In Fig.~\ref{fig:eval}a and Fig.~\ref{fig:eval}c,
the number of antennas is fixed to $\Mant=1024$ and the last case represents fully-digital beamforming ($\MRF=\Mant$) while
all the others refer to hybrid partially-connected beamforming. The corresponding \gls{pa} consumption 
(not depending on $\MRF$) is depicted on the rightmost bar of Fig.~\ref{fig:eval}b. In Fig.~\ref{fig:eval}a
and Fig.~\ref{fig:eval}b, we split the consumption of each
\gls{bs} component in load dependent (lighter bars) and load independent (darker bars). 
The load-dependent terms are the ones that depend on
$\overline{\subupp{x}{DL}}$ and $\overline{\subupp{x}{UL}}$ in~(\ref{eq:P_idig}),~(\ref{eq:P_iana}), and~(\ref{eq:P_pa}),
and vice versa for the load-independent terms.
We plot the full range of $\overline{\subupp{x}{DL}}+\overline{\subupp{x}{UL}}$, going from $0$ to $2$
and indicated with an arrow in one example. We also highlight more realistic loads,
i.e., $\overline{\subupp{x}{DL}}\in[0,0.3]$ and $\overline{\subupp{x}{UL}}\in[0,0.3]$.
At full loads, the total \gls{bs} consumption ranges from $740$ W (when $\MRF=16$) to $1658$ W (when $\MRF=1024$),
while it varies from $300$ W (when $\MRF=16$) to $1044$ W (when $\MRF=1024$) at
$\overline{\subupp{x}{UL}}=\overline{\subupp{x}{DL}}=0.3$.
We observe that, at full loads, the \gls{pa} consumption is the largest among the \gls{bs} components.
At more realistic loads, however, the \gls{pa} consumption becomes comparable to the sum of digital and analog consumption 
from $\MRF=256$. The analog and digital processing have larger load-independent consumption compared to the 
load-dependent one, while it is the opposite for the \gls{pa}. Fig.~\ref{fig:eval}b shows that the consumption of $\Mant$
\glspl{pa} at load of $\overline{\subupp{x}{DL}}=0.3$ varies from $4$ W (for $\Mant=16$) to $262$ W (for $\Mant=1024$).

In Fig.~\ref{fig:eval}c, we plot the ergodic rates in \gls{dl} and \gls{ul} for 
the full range of $\overline{\subupp{x}{DL}}$ and $\overline{\subupp{x}{UL}}$
(corresponding to the height of each bar), and we highlight the values at loads of $30\%$. 
We point out that for small numbers of \gls{rf} chains,
the rates do not drastically drop because there are $\Mant/\MRF=1024/\MRF$ subarrays that perform analog beamforming.
The \gls{dl} sum rate reaches more than $13$ Gbit/s and more than $2$ Gbit/s in \gls{ul} for 
the fully-digital beamforming at full load. We observe that hybrid beamforming with $128$ or more \gls{rf} chains
provides rates larger than $50\%$ of the ones achieved in both \gls{dl} and \gls{ul} by the fully-digital counterpart.
The \gls{ee}, computed as $\big(\subupp{R}{DL}+\subupp{R}{UL}\big)/\Pcons$ and depending on the loads, 
is equal to $\{3.6, 6.5, 9.6, 11.6, 13.4, 11.9, 9.5\}$ Mbit/J for $\MRF=\{16,32,64,128,256,512,1024\}$
at full loads. At $\overline{\subupp{x}{DL}}=\overline{\subupp{x}{UL}}=0.3$, it is equal to
$\{2.6, 4.7, 6.8, 7.6, 8.0, 6.3, 4.5\}$ Mbit/J. The maximal \gls{ee} of $13.6$ Mbit/J is therefore achieved by 
the hybrid beamforming configuration with $256$ \gls{rf} chains at full loads. The same configuration also improves
the \gls{ee} by a factor $1.78\times$ at loads of $30\%$.

\section{Conclusion}
In this study, we proposed a power model for \gls{FR3} \glspl{bs} that quantifies the
consumption of digital, analog, \gls{pa}, and supply and cooling components by averaging over different working modes
in a time frame, from active to idle. The consumption of each digital subcomponent is adapted to the hardware architecture, 
being this \gls{fpga} or \gls{asic}. Up-to-date power consumption scalings for analog subcomponents and \gls{pa} 
at \gls{FR3} are provided. The analysis compares hybrid partially-connected and fully-digital beamforming, and is 
mainly focused on a scenario where the \gls{bs} has $1024$ antennas, $16$ to $1024$ \gls{rf} chains,
and uses $400$ MHz of bandwidth. We show that, at resource loads until $30\%$, the power consumed by digital and 
analog components is larger than the \gls{pa} consumption when $512$ or more \gls{rf} chains are utilized. 
When $\MRF=\Mant=1024$, the individual consumption of digital and analog processing exceeds the one of the \glspl{pa}
Notably, hybrid beamforming achieves up to $13.6$ Mbit/J of \gls{ee} and $1.78\times$ \gls{ee} improvement over 
fully-digital beamforming.

\balance
\bibliographystyle{IEEEbib}
\bibliography{IEEEabrv,refs}

\begin{thebibliography}{10}

\bibitem{Andrews_2024}
J.~G. Andrews, T.~E. Humphreys, and T.~Ji,
\newblock ``{6G} takes shape,''
\newblock {\em IEEE BITS Inf.~Theory Mag.}, vol. 4, no. 1, pp. 2--24, 2024.

\bibitem{Kang_2024}
S.~Kang, M.~Mezzavilla, S.~Rangan, A.~Madanayake, S.~B. Venkatakrishnan, G.~Hellbourg, et~al.,
\newblock ``Cellular wireless networks in the upper mid-band,''
\newblock {\em IEEE Open J.~Commun.~Soc.}, vol. 5, pp. 2058--2075, 2024.

\bibitem{Zhang_2025}
J.~Zhang, H.~Miao, P.~Tang, L.~Tian, and G.~Liu,
\newblock ``New mid-band for {6G}: Several considerations from the channel propagation characteristics perspective,''
\newblock {\em IEEE Commun.~Mag.}, vol. 63, no. 1, pp. 175--180, 2025.

\bibitem{Shakya_2024}
D.~Shakya, M.~Ying, T.~S. Rappaport, H.~Poddar, P.~Ma, Y.~Wang, et~al.,
\newblock ``Comprehensive {FR1(C)} and {FR3} lower and upper mid-band propagation and material penetration loss measurements and channel models in indoor environment for {5G} and {6G},''
\newblock {\em IEEE Open J.~Commun.~Soc.}, vol. 5, pp. 5192--5218, 2024.

\bibitem{Mezzavilla_2025}
M.~Mezzavilla, A.~Rasteh, M.~Zappe, E.~Zappe, A.~Dhananjay, and S.~Rangan,
\newblock ``{6G} prototyping in the upper mid-band (7--24 {GHz}),''
\newblock in {\em Proc.~IEEE Wireless Commun.~Netw. Conf.}, 2025, pp. 1--3.

\bibitem{Bjornson_2025}
E.~Björnson, F.~Kara, N.~Kolomvakis, A.~Kosasih, P.~Ramezani, and M.~B. Salman,
\newblock ``Enabling {6G} performance in the upper mid-band by transitioning from massive to gigantic {MIMO},''
\newblock {\em IEEE Open J.~Commun.~Soc.}, pp. 1--1, 2025.

\bibitem{Busch_2024}
A.~M. Busch, K.~Eger, and B.~Richerzhagen,
\newblock ``Comparison of power consumption models for {5G} cellular network base stations,''
\newblock in {\em Intelligent Distributed Computing XVI}, M.~K{\"o}hler-Bu{\ss}meier, W.~Renz, and J.~Sudeikat, Eds., Cham, Switzerland, 2024, pp. 199--212, Springer Nature Switz.

\bibitem{Golard_2024}
L.~Golard, Y.~Agram, F.~Rottenberg, F.~Quitin, D.~Bol, and J.~Louveaux,
\newblock ``A parametric power model of multi-band sub-6 {GHz} cellular base stations using on-site measurements,''
\newblock in {\em Proc.~IEEE Int.~Symp.~Pers.~Indoor Mob.~Radio Commun.}, 2024, pp. 1--7.

\bibitem{Desset_2020}
C.~Desset, P.~Wambacq, Y.~Zhang, M.~Ingels, and A.~Bourdoux,
\newblock ``A flexible power model for mm-wave and {THz} high-throughput communication systems,''
\newblock in {\em Proc.~IEEE Int.~Symp.~Pers.~Indoor Mob.~Radio Commun.}, 2020, pp. 1--6.

\bibitem{Ribeiro_2018}
L.~N. Ribeiro, S.~Schwarz, M.~Rupp, and A.~L.~F. de~Almeida,
\newblock ``Energy efficiency of {mmWave} massive {MIMO} precoding with low-resolution {DACs},''
\newblock {\em IEEE J.~Sel.~Top.~Signal Process.}, vol. 12, no. 2, pp. 298--312, 2018.

\bibitem{Lopez-Perez_2024}
D.~López-Pérez, N.~Piovesan, and G.~Geraci,
\newblock ``Capacity and power consumption of multi-layer {6G} networks using the upper mid-band,''
\newblock in {\em Proc.~IEEE Int.~Conf.~Commun.}, 2025.

\bibitem{Peschiera_2025}
E.~Peschiera, Y.~Agram, F.~Quitin, L.~Van der Perre, and F.~Rottenberg,
\newblock ``On optimizing time-, space- and power-domain energy-saving techniques for sub-6 {GHz} base stations,''
\newblock {\em arXiv preprint arXiv:2505.15445}, 2025.

\bibitem{zcu216}
{AMD},
\newblock ``{Zynq UltraScale+ RFSoC ZCU216 Evaluation Kit},'' Evaluation board user guide (UG1390), Dec. 2023.

\bibitem{Boutros_2021}
A.~Boutros and V.~Betz,
\newblock ``{FPGA} architecture: Principles and progression,''
\newblock {\em IEEE Circuits Syst. Mag.}, vol. 21, no. 2, pp. 4--29, 2021.

\bibitem{dahlman20205g}
Erik Dahlman, Stefan Parkvall, and Johan Sköld,
\newblock {\em {5G NR: The Next Generation Wireless Access Technology}},
\newblock Academic Press, New York, NY, USA, 2nd edition, 2020.

\bibitem{Cao_2022}
W.~Cao, S.~Wang, P.~N. Landin, C.~Fager, and T.~Eriksson,
\newblock ``Complexity optimized digital predistortion model of {RF} power amplifiers,''
\newblock {\em IEEE Trans.~Microw. Theory Tech.}, vol. 70, no. 3, pp. 1490--1499, 2022.

\bibitem{Mezzavilla_2024}
M.~Mezzavilla, A.~Dhananjay, M.~Zappe, and S.~Rangan,
\newblock ``A frequency hopping software-defined radio platform for communications and sensing in the upper mid-band,''
\newblock in {\em Proc.~IEEE Int.~Workshop Signal Process.~Adv.~Wireless Commun.}, 2024, pp. 611--615.

\bibitem{Shuguang_2005}
S.~Cui, A.~J. Goldsmith, and A.~Bahai,
\newblock ``Energy-constrained modulation optimization,''
\newblock {\em IEEE Trans.~Wireless Commun.}, vol. 4, no. 5, pp. 2349--2360, 2005.

\bibitem{Sihao_2025}
S.~Chen, C.~Huang, L.~Sun, Y.~Liu, W.~Tang, J.~Fan, et~al.,
\newblock ``A 28-nm 8-bit {16-GS/s DAC} with {\textgreater}60 {dBc}/{\textgreater}40 {dBc} {SFDR} up to 2.3 {GHz/5.4 GHz} using 4-channel {NRZ}-output-overlapped time-interleaving,''
\newblock {\em IEEE Trans.~Circuits Syst.~II, Exp.~Briefs}, vol. 72, no. 2, pp. 374--378, 2025.

\bibitem{adc_survey}
B.~Murmann,
\newblock ``{ADC Performance Survey 1997-2025},'' [Online]. Available: https://github.com/bmurmann/ADC-survey.

\bibitem{Markulic_2024}
N.~Markulić, J.~Nguyen, J.~L. Benites, E.~Martens, and J.~Craninckx,
\newblock ``A {10GS/s} hierarchical time-interleaved {ADC} for {RF}-sampling applications,''
\newblock in {\em Proc.~IEEE Symp.~VLSI Technol.~Circuits}, 2024, pp. 1--2.

\bibitem{Helal_2025}
M.~Helal, A.~Helaly, and G.~M. Rebeiz,
\newblock ``A 10--36-{GHz I/Q} mixer-first {RX} using current-mode coupler cascade in 16-nm {FinFET},''
\newblock {\em IEEE Microw.~Wireless Technol.~Lett.}, pp. 1--4, 2025.

\bibitem{Abbasi_2025}
H.~Abbasi, A.~Slater, F.~Beheshti, S.~Poolakkal, and S.~Gupta,
\newblock ``A four-element true-time-delay slice-based receiver array for {FR3} upper mid-band wireless,''
\newblock {\em IEEE J.~Solid-State Circuits}, pp. 1--15, 2025.

\bibitem{Afifi_2025}
A.~Afifi and G.~M. Rebeiz,
\newblock ``A multi-band 5.3--18 {GHz LNA for FR3} bands using hybrid switching with sub-2 {dB NF} in {FDSOI},''
\newblock {\em IEEE J.~Solid-State Circuits}, pp. 1--11, 2025.

\bibitem{ADMV8818}
{Analog Devices},
\newblock ``{ADMV8818: 2 {GHz to 18 GHz}, Digitally Tunable, High-Pass and Low-Pass Filter},'' Data Sheet (Rev. B), 2021.

\bibitem{ADF41510}
{Analog Devices},
\newblock ``{ADF41510: 10 {GHz}, Integer-{N}/Fractional-{N} {PLL} Synthesizer},'' Data Sheet (Rev. A), 2024.

\bibitem{Cheng_2014}
J.-F. Cheng, H.~Koorapaty, P.~Frenger, D.~Larsson, and S.~Falahati,
\newblock ``Energy efficiency performance of {LTE} dynamic base station downlink {DTX} operation,''
\newblock in {\em Proc. IEEE 79th Veh. Technol. Conf.}, 2014, pp. 1--5.

\bibitem{Wang_2024}
H.~Wang, M.~Eleraky, B.~Abdelaziz, et~al.,
\newblock ``Power amplifiers performance survey 2000--present,'' 2025,
\newblock [Online]. Available: https://ideas.ethz.ch/research/surveys/pa-survey.html.

\bibitem{Pecile_2025}
D.~Pecile, A.~Gambarucci, S.~Kokorovic, and A.~Bevilacqua,
\newblock ``Analysis and design of a {SiGe BiCMOS PA for 6G FR3} band with {29-dBm P\textsubscript{PAE}} and 40.1\% {PAE},''
\newblock {\em IEEE Trans.~Microw.~Theory Tech.}, pp. 1--12, 2025.

\bibitem{sionna}
J.~Hoydis, S.~Cammerer, F.~{Ait Aoudia}, M.~Nimier-David, L.~Maggi, G.~Marcus, et~al.,
\newblock ``Sionna,'' 2022,
\newblock https://nvlabs.github.io/sionna/.

\end{thebibliography}

\end{document}